\documentclass[12pt]{article}
\usepackage[dvips]{graphicx}

\newcommand{\ie}{{\it i.e.}}

\renewcommand{\bar}[1]{\overline{#1}}
\newcommand{\half}  {\frac{1}{2}}

\newcommand{\qu}{{\rm q}}

\newcommand{\ieps}{i\varepsilon}

\newcommand{\order}[1]{${ O}\left(#1 \right)$}

\newcommand{\beq}{\begin{equation}}
\newcommand{\eeq}{\end{equation}}
\newcommand{\beqa}{\begin{eqnarray}}
\newcommand{\eeqa}{\end{eqnarray}}
\newcommand{\VEV}[1]{\left\langle{#1}\right\rangle}
\newcommand{\ket}[1]{\vert{#1}\rangle}

\begin {document}
\begin{flushright}
{\small
SLAC--PUB--9055\\
November 2001\\}
\end{flushright}
 \vfill
\begin{center}
{{\bf\Large The Heisenberg Matrix Formulation \\[1ex] of Quantum
Field Theory}\footnote{Work supported by Department of Energy contract
DE--AC03--76SF00515.}}

\vskip 1truecm
{\large\bf Stanley J. Brodsky}
\vskip 5truemm
{Stanford Linear Accelerator Center \\
Stanford University, Stanford, California 94309 \\
{\tt  sjbth@slac.stanford.edu}}
\end{center}\vfill

\begin{center}
Presented at the  \\
 Symposium On 100 Years Werner Heisenberg: Works And Impact
\\
  Bamberg, Germany\\
26-30 September 2001 \\
\end{center}
\vfill

\newpage

\noindent {\bf Abstract:}

Heisenberg's matrix formulation of quantum mechanics can be
generalized to relativistic systems by evolving in light-front time
$\tau = t+z/c.$ The spectrum and wavefunctions of bound states,
such as hadrons in quantum chromodynamics, can be obtained from
matrix diagonalization of the light-front Hamiltonian on a finite
dimensional light-front Fock basis defined using periodic boundary
conditions in $x^-$ and $x_\perp$.  This method, discretized
light-cone quantization (DLCQ), preserves the frame-independence
of the front form even at finite resolution and particle number.
Light-front quantization can also be used in the Hamiltonian form
to construct an event generator for high energy physics reactions
at the amplitude level.  The light-front partition function, summed
over exponentially-weighted light-front energies, has simple boost
properties which may be useful for studies in heavy ion
collisions.  I also review recent work which shows that the
structure functions measured in deep inelastic lepton scattering
are affected by final-state rescattering, thus modifying their
connection to light-front probability distributions.  In
particular, the shadowing of nuclear structure functions is due to
destructive interference effects from leading-twist diffraction of
the virtual photon, physics not included in the nuclear light-front
wavefunctions.
 \vskip 1truecm

\section{Introduction}

One of the challenges of relativistic quantum field theory is to
compute the wavefunctions of bound states, such as the amplitudes
which determine the quark and gluon substructure of hadrons in
quantum chromodynamics.  However, any extension of the
Heisenberg-Schr\"odinger formulation of quantum mechanics $H
\ket{\psi} = i {\partial \over \partial t} \ket{\psi} = E
\ket{\psi}$ to the relativistic domain has to confront seemingly
intractable hurdles: (1) quantum fluctuations preclude finite
particle-number wavefunction representations; (2) the charged
particles arising from the quantum fluctuations of the vacuum
contribute to the matrix element of currents--- thus knowledge of
the wavefunctions alone is insufficient to determine observables;
and (3) the boost of an equal-time wavefunction from one Lorentz
frame to another not only changes particle number, but is as
complicated a dynamical problem as solving for the wavefunction
itself.

In 1949, Dirac~\cite{Dirac:cp} made the remarkable observation
that ordinary ``instant" time $t$ is not the only possible
evolution parameter. In fact,  evolution in ``light-front" time
$\tau = t + z/c = x^+$ has extraordinary advantages for
relativistic systems, stemming from the fact that a subset of the
Lorentz boost operators becomes purely kinematical.  In fact, the
Fock-state representation of bound states defined at equal
light-front time, \ie, along the light-front, provides
wavefunctions of fixed particle number which are independent of
the eigenstate's four-momentum $P^\mu.$ Furthermore, quantum
fluctuations of the vacuum are absent if one uses light-front time
to quantize the system, so that matrix elements such as the
electromagnetic form factors only depend on the currents of the
constituents described by the light-front wavefunctions.

In Dirac's ``Front Form", the generator of light-front time
translations is $P^- = i{ \partial\over \partial \tau}.$ Boundary
conditions are set on the transverse plane labeled by $x_\perp$
and $x^- = z-ct.$ See Fig.~{\ref{fig:bir2}.
Given the Lagrangian of a quantum field theory,
$P^-$ can be constructed as an operator on the Fock basis, the
eigenstates of the free theory.
(This method is also called ``light-cone" quantization in the literature.)
Since each  particle in the Fock
basis is on its mass shell, $k^- \equiv k^0-k^3 = {k^2_\perp + m^2
\over k^+},$ and its energy $k^0 =\half ( k^+ + k^-) $ is
positive, only particles with positive momenta $k^+ \equiv k^0 +
k^3 \ge 0$ can occur in the Fock basis. Since the total plus
momentum $P^+ = \sum_n k^+_n$ is conserved, the light-front vacuum
cannot have any particle content.   The operator $H_{LC} = P^+ P^-
- P^2_\perp,$ the ``light-front Hamiltonian", is frame-independent.

\begin{figure}
\includegraphics[width=6in]{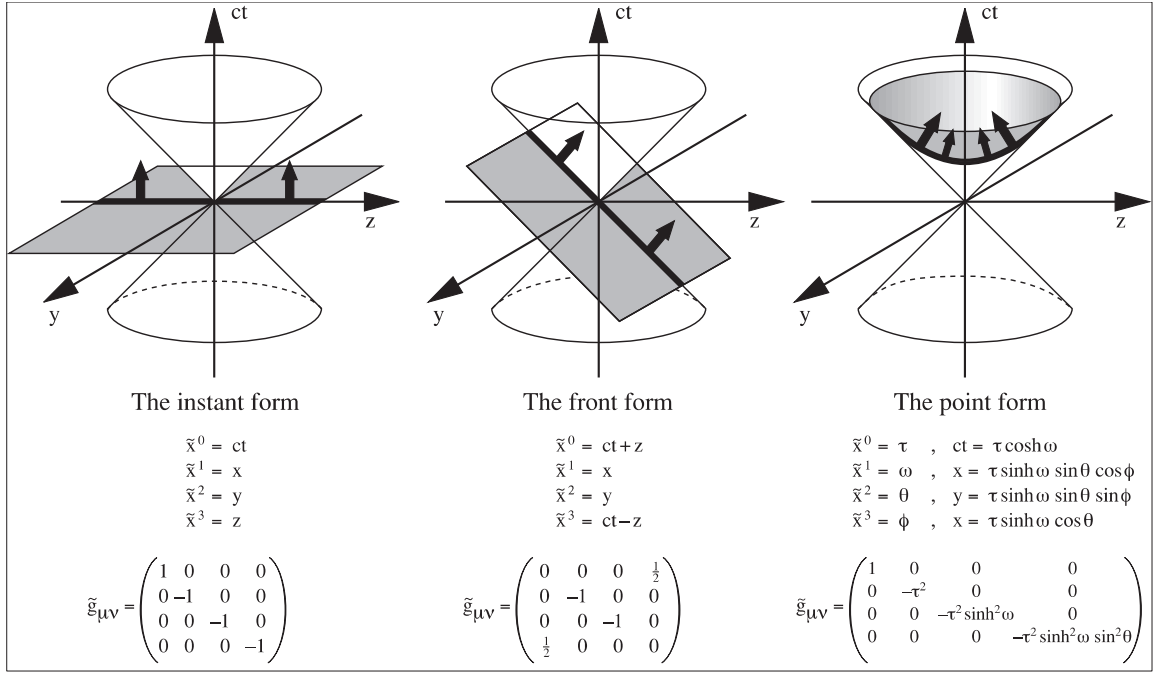}
\caption{\label{fig:bir2}
    Dirac's three forms of Hamiltonian dynamics.
From Ref. \cite{Brodsky:1997de}. }
\end{figure}

The Heisenberg equation on the light-front is
\begin{equation}
H_{LC} \ket{\Psi} = M^2 \ket{\Psi}\ .
\end{equation}
This can in principle be solved by diagonalizing the matrix
$\VEV{n|H_{LC}|m}$ on the free Fock basis:~\cite{Brodsky:1997de}
\begin{equation}
\sum_m
\VEV{n|H_{LC}|m}\VEV{m|\psi} = M^2 \VEV{n|\Psi}\ .
\end{equation}
The eigenvalues $\{M^2\}$ of $H_{LC}=H^{0}_{LC} + V_{LC}$ give the
squared invariant masses of the bound and continuum spectrum of
the theory.  For example, the light-cone gauge
interaction terms of QCD which are important for a meson
are illustrated in Fig.~\ref{fig-kyf-3}.
The projections $\{\VEV{n|\Psi}\}$ of the eigensolution
on the $n$-particle Fock states provide the light-front wavefunctions.
Thus solving a quantum field theory is equivalent to solving
a coupled many-body quantum mechanical problem:
\begin{equation}
\left[M^2 - \sum_{i=1}^n{m_{\perp i}^2\over x_i}\right] \psi_n =
\sum_{n'}\int \VEV{n|V_{LC}|n'} \psi_{n'}
\end{equation}
where the
convolution and sum is understood over the Fock number, transverse
momenta, plus momenta, and helicity of the intermediate states.
Here $m^2_\perp = m^2 + k^2_\perp.$

\begin{figure}
\centerline{\includegraphics[width=6in]{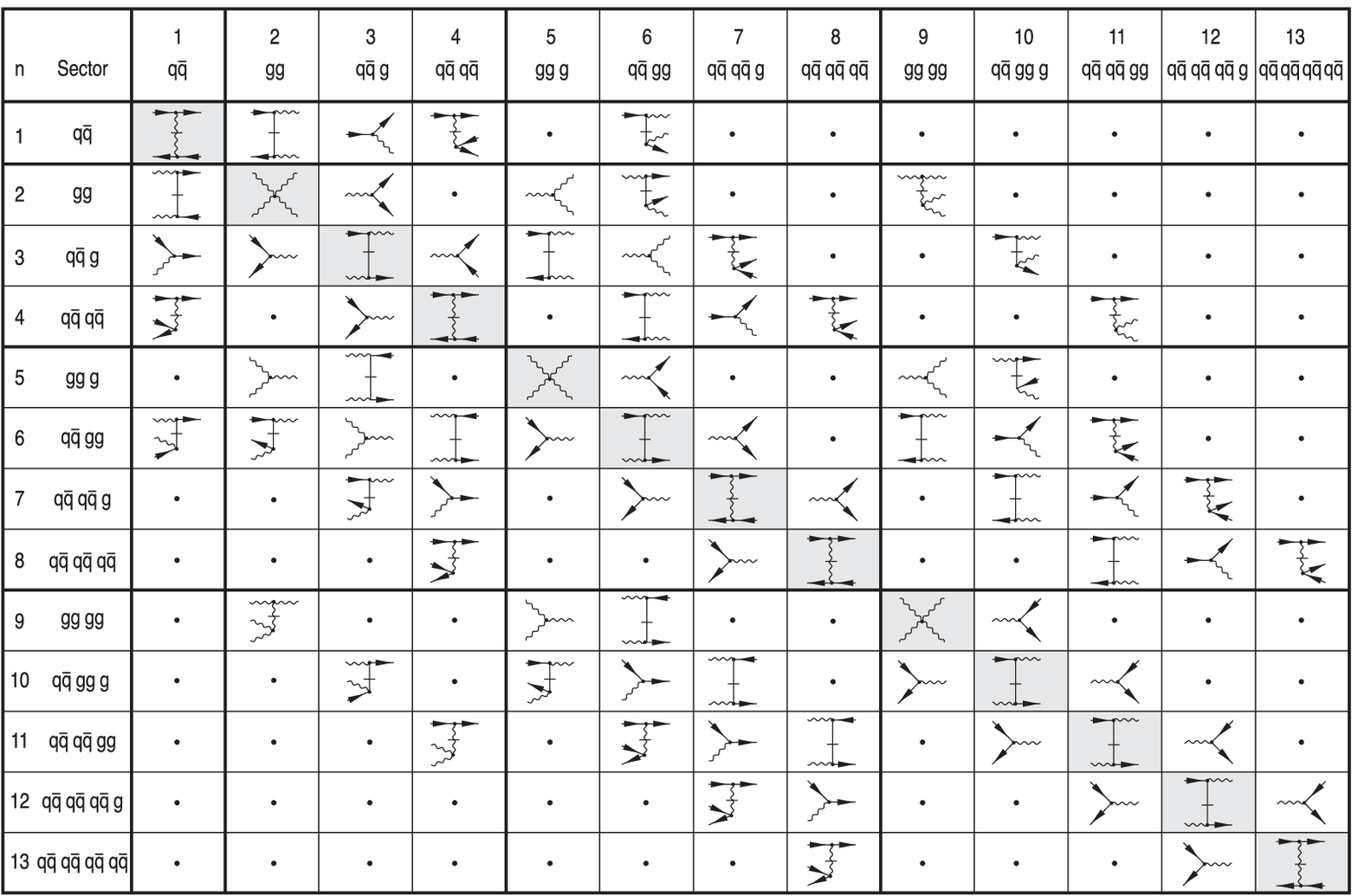}}
\caption{\label{fig-kyf-3} The front-form matrix of QCD
interactions in light-cone gauge. Up to eight constituents in a
meson are shown. From Ref. \cite{Brodsky:1997de} and H. C. Pauli.}
         \end{figure}


In QCD, the wavefunction of a hadron describes its
composition in terms of the momenta and spin projections of quark and
gluon constituents.  For example, the
eigensolution of a negatively-charged meson QCD,  projected on its
color-singlet $B = 0$, $Q = -1$, $J_z = 0$ eigenstates $\{\ket{n} \}$ of
the free Hamiltonian $ H^{QCD}_{LC}(g = 0)$ at fixed $\tau = t-z/c$ has
the expansion:
\begin{eqnarray}
\left\vert \Psi_M; P^+, {\vec P_\perp}, \lambda \right> &=& \sum_{n
\ge 2,\lambda_i} \int \Pi^{n}_{i=1} {d^2k_{\perp i} dx_i \over
\sqrt{x_i} 16 \pi^3}
 16 \pi^3 \delta\left(1- \sum^n_j x_j\right) \delta^{(2)}
\left(\sum^n_\ell \vec k_{\perp \ell}\right) \nonumber \\[1ex]
&&\left\vert n; x_i P^+, x_i {\vec P_\perp} + {\vec k_{\perp i}},
\lambda_i\right
> \psi_{n/M}(x_i,{\vec k_{\perp i}},\lambda_i)
 .
\end{eqnarray}
The set of light-front Fock state wavefunctions $\{\psi_{n/M}\}$
represent the ensemble of quark and gluon states possible when the
meson is intercepted at the light-front.  The light-front momentum
fractions $x_i = k^+_i/P^+_\pi = (k^0 + k^z_i)/(P^0+P^z)$ with
$\sum^n_{i=1} x_i = 1$ and ${\vec k_{\perp i}}$ with $\sum^n_{i=1}
{\vec k_{\perp i}} = {\vec 0_\perp}$ represent the relative
momentum coordinates of the QCD constituents and are independent
of the total momentum of the state.  The actual physical
transverse momenta are ${\vec p_{\perp i}} = x_i {\vec P_\perp} +
{\vec k_{\perp i}}.$ The $\lambda_i$ label the light-front spin
$S^z$ projections of the quarks and gluons along the quantization
$z$ direction.  The spinors of the light-front formalism
automatically incorporate the Melosh-Wigner rotation.  The
physical gluon polarization vectors $\epsilon^\mu(k,\ \lambda =
\pm 1)$ are specified in light-cone gauge by the conditions $k
\cdot \epsilon = 0,\ \eta \cdot \epsilon = \epsilon^+ = 0.$ The
parton degrees of freedom are thus all physical; there are no
ghost or negative metric states.  A detailed derivation of
light-front quantization of non-Abelian gauge theory in light-cone
gauge is given in Ref.~\cite{Srivastava:2000cf}. Explicit examples
of light-front wavefunctions in QED are given in Ref.
\cite{Brodsky:2001ii}.

Angular momentum has simplifying features in the light-front formalism
since the projection $J_z$ is kinematical and
conserved.  Each light-front Fock wavefunction satisfies the angular
momentum sum rule: $ J^z = \sum^n_{i=1} S^z_i + \sum^{n-1}_{j=1}
l^z_j \ . $ The sum over $S^z_i$ represents the contribution of
the intrinsic spins of the $n$ Fock state constituents.  The sum
over orbital angular momenta
\begin{equation}
l^z_j = -{\mathrm i} \left(k^1_j\frac{\partial}{\partial k^2_j}
-k^2_j\frac{\partial}{\partial k^1_j}\right) \end{equation}
 derives from
the $n-1$ relative momenta.  This excludes the contribution to the
orbital angular momentum due to the motion of the center of mass,
which is not an intrinsic property of the
hadron~\cite{Brodsky:2001ii}. The numerator structure of the
light-front wavefunctions is in large part determined by the
angular momentum constraints.

The most important feature of light-front Fock wavefunctions
$\psi_{n/p}(x_i,\vec k_{\perp i},\lambda_i)$ is the fact they are
Lorentz invariant functions of the relative coordinates,
independent of the bound state's physical momentum $P^+ = P^0 +
P^z$, and $P_\perp$~\cite{Lepage:1980fj}. The light-front
wavefunctions represent the ensembles of states possible when the
hadron is intercepted by a light-front at fixed $\tau = t+z/c.$
The light-front representation thus provide a frame-independent,
quantum-mechanical representation of a hadron at the amplitude
level, capable of encoding its multi-quark, hidden-color and gluon
momentum, helicity, and flavor correlations in the form of
universal process-independent hadron wavefunctions.

If one imposes periodic boundary conditions in $x^- = t + z/c$,
then the plus momenta become discrete: $k^+_i = {2\pi \over L}
n_i, P^+ = {2\pi\over L} K$, where $\sum_i n_i =
K$~\cite{Maskawa:1975ky,Pauli:1985pv}.  For a given ``harmonic
resolution" $K$, there are only a finite number of ways positive
integers $n_i$ can sum to a positive integer $K$.  Thus at a given
$K$, the dimension of the resulting light-front Fock state
representation of the bound state is rendered finite without
violating Lorentz invariance.  The eigensolutions of a quantum
field theory, both the bound states and continuum solutions, can
then be found by numerically diagonalizing a frame-independent
light-front Hamiltonian $H_{LC}$ on a finite and discrete
momentum-space Fock basis.  Solving a quantum field theory at
fixed light-front time $\tau$ thus can be formulated as a
relativistic extension of Heisenberg's matrix mechanics.  The
continuum limit is reached for $K \to \infty.$ This formulation of
the non-perturbative light-front quantization problem is called
``discretized light-front quantization" (DLCQ)~\cite{Pauli:1985pv}.
Lattice gauge theory has also been used to calculate the pion
light-front wavefunction~\cite{Abada:2001if}.

The DLCQ method has been used extensively for solving one-space
and one-time theories~\cite{Brodsky:1997de}, including
applications to supersymmetric quantum field
theories~\cite{Matsumura:1995kw} and specific tests of the
Maldacena conjecture~\cite{Hiller:2001mh}. There has been progress
in systematically developing the computation and renormalization
methods needed to make DLCQ viable for QCD in physical spacetime.
For example, John Hiller, Gary McCartor, and
I~\cite{Brodsky:2001ja} have shown how DLCQ can be used to solve
3+1 theories despite the large numbers of degrees of freedom
needed to enumerate the Fock basis.  A key feature of our work is
the introduction of Pauli Villars fields to regulate the UV
divergences and perform renormalization while preserving the
frame-independence of the theory.  A recent application of DLCQ to
a 3+1 quantum field theory with Yukawa interactions is given in
Ref.~\cite{Brodsky:2001ja}. Representative plots of the one-boson
one-fermion light-front Fock wavefunction of the lowest mass
fermion solution of the Yukawa (3+1) theory  showing spin
correlations and the presence of non-zero orbital angular momentum
are shown in Fig.~\ref{fig:phi1000}.

There has also been important progress using the transverse
lattice, essentially a combination of DLCQ in 1+1 dimensions
together with a lattice in the transverse
dimensions~\cite{Bardeen:1979xx,Dalley:2001gj,Burkardt:2001dy}.
One can also define a truncated theory by eliminating the higher
Fock states in favor of an effective
potential~\cite{Pauli:2001vi}.  Spontaneous symmetry breaking and
other nonperturbative effects associated with the instant-time
vacuum are hidden in dynamical or constrained zero modes on the
light-front.  An introduction is given in
Refs.~\cite{McCartor:hj,Yamawaki:1998cy}.

\begin{figure}[htb]
{\includegraphics[width=3.2in]{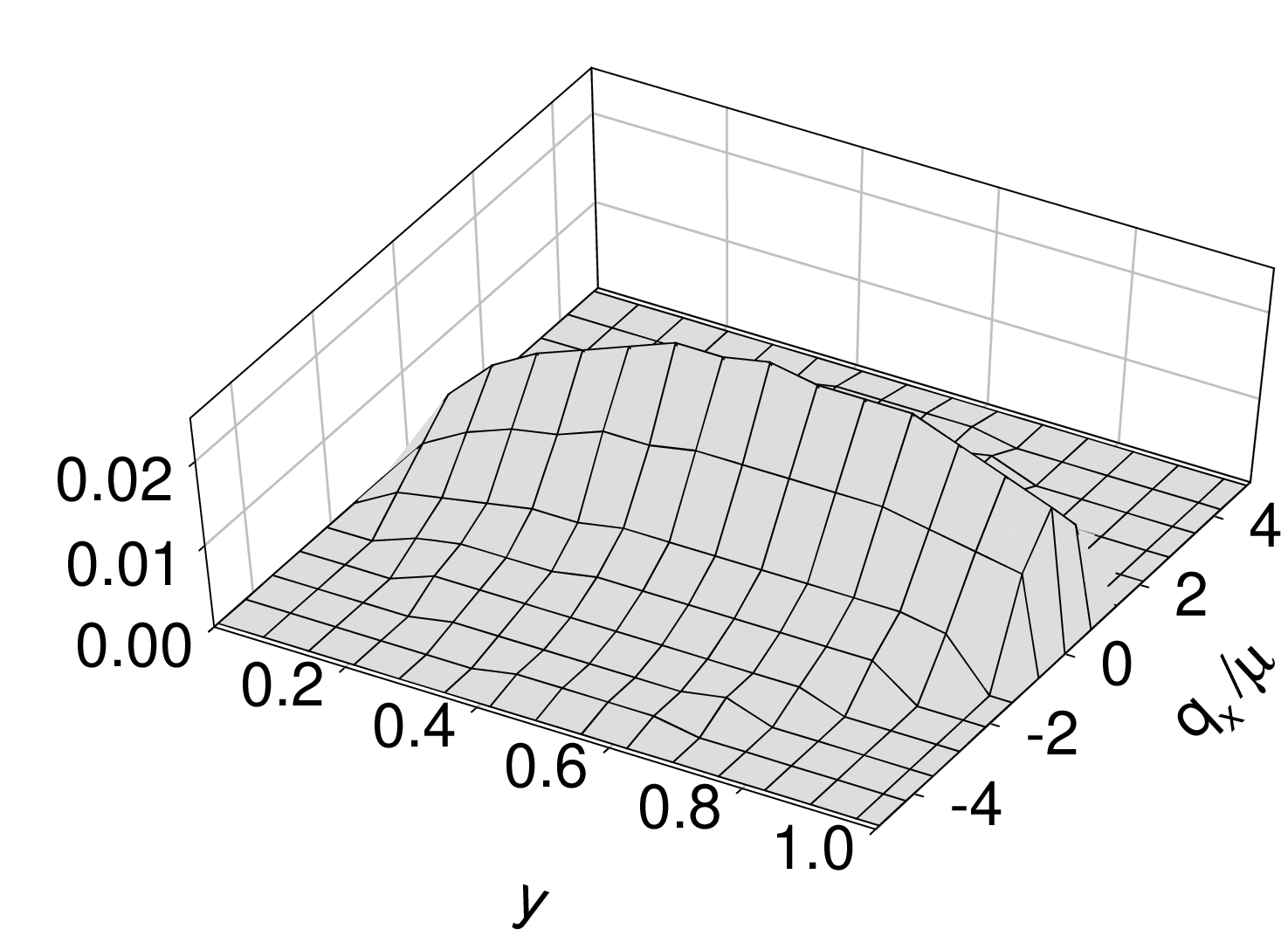}\hfill
\includegraphics[width=3.2in]{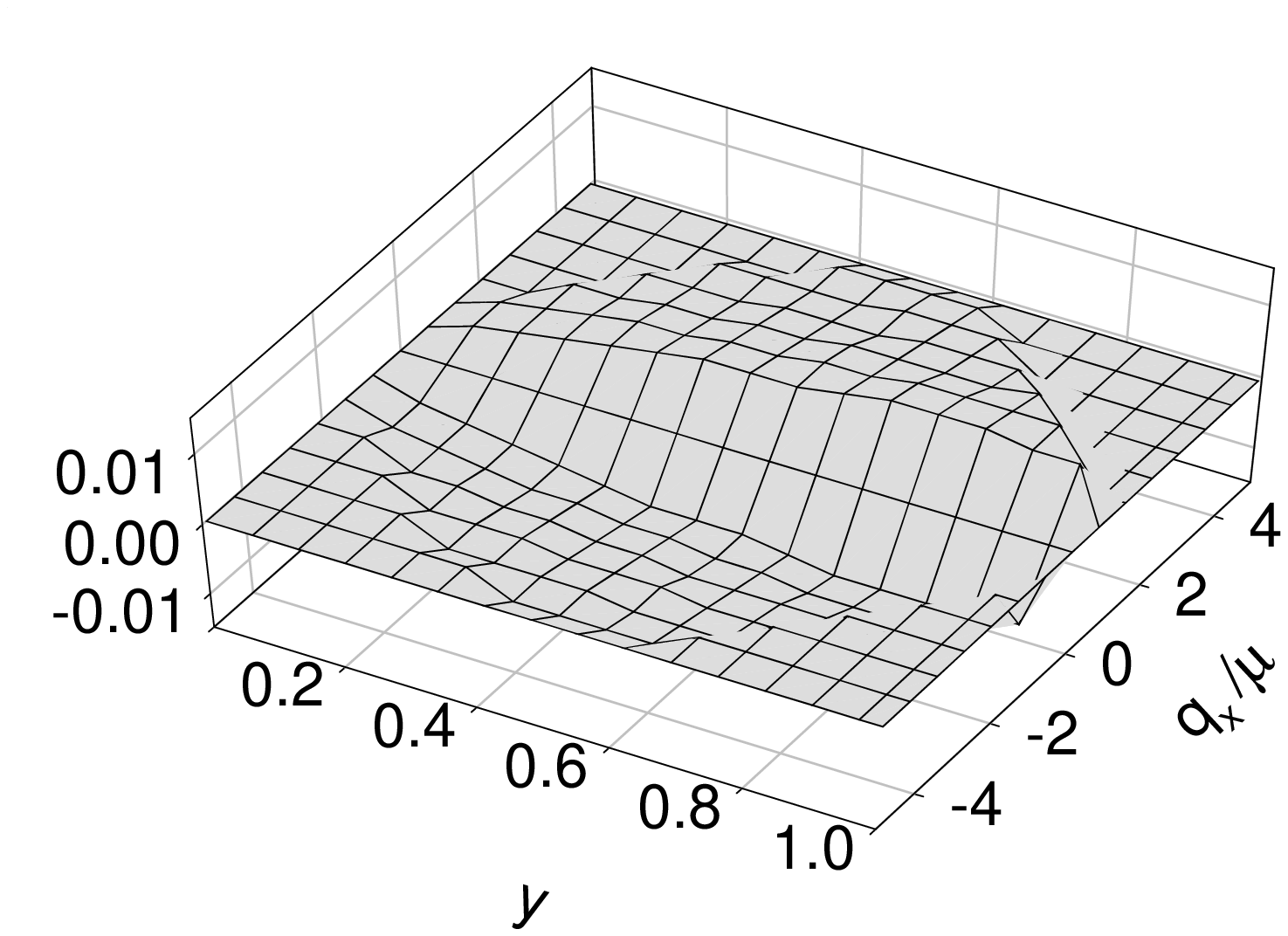}} \caption[*]{\label{fig:phi1000}
DLCQ results for the one-boson one-fermion wavefunction in a
fermion system with parallel  and antiparallel  fermion helicity,
as a function of longitudinal momentum fraction $y$ and one
transverse momentum component $q_x$ in the $q_y=0$ plane. The
parameter values for the DLCQ resolution are $K=29$, $N_\perp=7$.
Further details are given in Ref.~\cite{Brodsky:2001ja}. }
\end{figure}

\section{General Features of Light-front Wavefunctions}

The
maximum of a light-front wavefunction occurs when the invariant mass of
the partons is minimal; \ie, when all particles have equal rapidity
and are all at rest in the rest frame.  In fact, Dae Sung Hwang
and I \cite{BH2} have noted that one can rewrite the wavefunction
in the form: \begin{equation} \psi_n= {\Gamma_n\over M^2
[\sum_{i=1}^n {(x_i-{\widehat x}_i)^2\over x_i} + \delta^2]}
\end{equation}
where $x_i = {\widehat x}_i\equiv{m_{\perp i}/ \sum_{i=1}^n
m_{\perp i}}$ is the condition for minimal rapidity differences of
the constituents.  The key parameter is $
M^2-\sum_{i=1}^n{m_{\perp i}^2/ {\widehat x}_i}\equiv
-M^2\delta^2.$ One can interpret $\delta^2 \simeq 2 \epsilon /
M $ where $ \epsilon = \sum_{i=1}^n m_{\perp i}-M $ is the
effective binding energy.  This form shows that the wavefunction is
a quadratic form around its maximum, and that the width of the
distribution in $(x_i - \widehat x_i)^2$ (where the wavefunction
falls to half of its maximum) is controlled by $x_i \delta^2$ and
the transverse momenta $k_{\perp_i}$.  Note also that the heaviest
particles tend to have the largest $\widehat x_i,$ and thus the
largest momentum fraction of the particles in the Fock state, a
feature familiar from the intrinsic charm model.  For example, the
$b$ quark has the largest momentum fraction at small $k_\perp$ in
the $B$ meson's valence light-front wavefunction, but the
distribution spreads out to an asymptotically symmetric
distribution around $x_b \sim 1/2$ when $k_\perp \gg m^2_b.$

The fall-off the light-front wavefunctions at large $k_\perp$ and
$x \to 1$ is dictated by QCD perturbation theory since the state
is far-off the light-front energy shell.  This leads to counting
rule behavior for the quark and gluon distributions at $x \to 1$.
Notice that $x\to 1$ corresponds to $k^z \to -\infty$ for any
constituent with nonzero mass or transverse momentum.

The above discussion suggests that an approximate form for the
hadron light-front wavefunctions could be constructed through
variational principles and by minimizing the expectation value of
$H^{QCD}_{LC}.$

\section{A Light-front Event Amplitude Generator}

The light-front formalism can be used as an
``event amplitude generator" for high energy physics reactions where each
particle's final state is completely labeled in momentum, helicity,
and phase.  The application of the light-front time evolution operator
$P^-$ to an initial state systematically generates the tree and
virtual loop graphs of the
$T$-matrix in light-front time-ordered perturbation theory in light-front
gauge.  The loop integrals only involve integrations over the
momenta of physical quanta and physical phase space
$\prod d^2k_{\perp i} d k^+_i$.  Renormalized amplitudes can be explicitly
constructed by subtracting from the divergent loops amplitudes with
nearly identical integrands corresponding to the contribution of the
relevant mass and coupling counter terms (the ``alternating denominator
method")~\cite{Brodsky:1973kb}.  The natural renormalization scheme to
use for defining the coupling in the event amplitude generator is a
physical effective charge such as the pinch scheme~\cite{Cornwall:1989gv}.
The argument of the coupling is then
unambiguous~\cite{Brodsky:1994eh}.  The DLCQ boundary conditions can be
used to discretize the phase space and limit the number of contributing
intermediate states without violating Lorentz invariance.
Since one avoids dimensional regularization and nonphysical ghost degrees
of freedom, this method of generating events at the amplitude level
could provide a simple but powerful tool for simulating events both
in QCD and the Standard Model.

\section{The Light-front Partition Function}

In the usual treatment of classical thermodynamics, one considers
an ensemble of particles $n = 1, 2, \ldots N$ which have energies
$\{E_n\}$ at a given ``instant" time $t$. The partition function
is defined as $Z = \sum_n \exp-{E_n\over kT}.$ Similarly, in
quantum mechanics, one defines a quantum-statistical partition
function as $Z = tr \exp{-\beta H}$ which sums over the
exponentiated-weighted energy eigenvalues of the system.

In the case of relativistic systems, it is natural to characterize
the system at a given light-front time $\tau = t +z/c$; {\em i.e.},
one determines the state of each particle in the ensemble as its
encounters the light-front. Thus we can define a light-front
partition function
$$Z_{LC} = \sum_n \exp -{p^-_n\over kT_{LC}}$$
by summing over the particles' light-front energies $p^- = p^0 -
p^z = {p^2_\perp + m^2 \over p^+}$.  The total momentum is $P^+ =
\sum p^+_n,$ $ \vec P_\perp = \sum_n \vec p_{\perp n}$, and the
total mass is defined from $P^+P^--P^2_\perp=M^2$.  The product
${M \over P^-} T_{LC}$ is boost invariant.  In the center of mass
frame where $\vec P =0$ and thus $P^+ = P^- = M$.  It is also
possible to consistently impose boundary conditions at fixed $x^-
= z - ct$ and $x_\perp$, as in DLCQ.  The momenta $p^+_n, \vec
p_{\perp n}$ then become discrete. The corresponding light-front
quantum-statistical partition function is $Z = tr \exp{-\beta_{LC}
H_{LC}}$ where $H_{LC} = i {\partial\over
\partial \tau}$ is the light-front Hamiltonian.

For non-relativistic systems the light-front partition function reduces to
the standard definition.  However, the light-front
partition function should be advantageous for analyzing relativistic systems
such as heavy ion collisions, since, like true rapidity, $y = \ln
{p^+\over P^+},$ light-front variables have simple behavior under
Lorentz boosts.  The light-front formalism also takes into account the
point that a phase transition does not occur simultaneously in $t$, but
propagates through the system with a finite wave velocity.

\section{Light-front Wavefunctions and QCD Phenomenology}

There have been extensive applications of light-front wavefunctions
to QCD phenomenology~\cite{Brodsky:2001dx}; for example, form
factors~\cite{Brodsky:1980zm} and the handbag contribution to
deeply virtual Compton scattering $\gamma^* p \to \gamma p$ can be
expressed as overlaps of the light-front
wavefunctions~\cite{Brodsky:2000xy,Diehl:2000xz}; quark and gluon
distributions are light-front wavefunction probabilities.  The
distributions measured in the diffractive dissociation of hadrons
are computed from transverse derivatives of the light-front
wavefunctions. Progress in measuring the basic parameters of
electroweak interactions and $CP$ violation will require a
quantitative understanding of the dynamics and phase structure of
$B$ decays at the amplitude level. The light-front Fock
representation is specially advantageous in the study of exclusive
$B$ decays.  For example, Dae Sung Hwang~\cite{Brodsky:1999hn} and
I have derived an exact frame-independent representation of decay
matrix elements such as $B \to D \ell \bar \nu$ from the overlap
of $n' = n$ parton-number conserving wavefunctions and the overlap
of wavefunctions with $n' = n-2$ from the annihilation of a
quark-antiquark pair in the initial wavefunction.

One can also express the matrix elements of the energy momentum
tensor as overlap integrals of the light-front
wavefunctions~\cite{Brodsky:2001ii}.  An important consistency
check of any relativistic formalism is to verify the vanishing of
the anomalous gravito-magnetic moment $B(0)$, the spin-flip matrix
element of the graviton coupling and analog of the anomalous
magnetic moment $F_2(0)$. For example, at one-loop order in QED,
$B_f(0) = {\alpha \over 3 \pi}$ for the electron when the graviton
interacts with the fermion line, and $B_\gamma(0) = -{\alpha \over
3 \pi}$ when the graviton interacts with the exchanged photon.
The vanishing of $B(0)$ can be shown to be exact for bound or
elementary systems in the light-front
formalism~\cite{Brodsky:2001ii}, in agreement with the equivalence
principle~\cite{Okun,Kob70,Teryaev}.

\section{Structure Functions are Not Parton Distributions}

The quark and gluon distributions of hadrons can be defined from the
probability measures of the light-front wavefunctions.  For example, the
quark distribution in a hadron $H$ is
\begin{equation}
{ P}_{\qu/H}(x_{Bj},Q^2)= \sum_n \int^{k_{i\perp}^2<Q^2}\left[
\prod_i\, dx_i\, d^2k_{\perp i}\right] |\psi_{n/H}(x_i,k_{\perp i})|^2
\sum_{j=q} \delta(x_{Bj}-x_j).
\end{equation}
It has been conventional to identify the leading-twist structure
functions $F_i(x,Q^2)$ measured in deep inelastic lepton
scattering with the light-front probability distributions.  For
example, in the parton model, $F_2(x,Q^2) = \sum_q e^2_q {
P}_{\qu/H}(x,Q^2).$ However, Paul Hoyer, Nils Marchal, Stephane
Peigne, Francesco Sannino, and I \cite{Brodsky:2001ue} have
recently shown that the leading-twist contribution to deep
inelastic scattering is affected by diffractive rescattering of a
quark in the target, a coherent effect which is not included in
the light-front wavefunctions, even in light-cone gauge.  The gluon
propagator in light-cone gauge $A^+=0$ is singular:
\begin{equation}
d_{LC}^{\mu\nu}(k) =
\frac{i}{k^2+\ieps}\left[-g^{\mu\nu}+\frac{n^\mu k^\nu+ k^\mu
n^\nu}{n\cdot k}\right] \label{lcprop}
\end{equation}
has a pole
at $k^+ \equiv n\cdot k = 0,$ which has to be defined by an
analytic prescription such as the Mandelstam-Liebbrandt
prescription~\cite{Leibbrandt:1987qv}.  In final-state scattering
involving on-shell intermediate states, the exchanged momentum
$k^+$ is of \order{1/\nu} in the target rest frame, which enhances
the second term of the light-cone gauge propagator.  This
enhancement allows rescattering to contribute at leading twist
even in LC gauge.

Thus diffractive contributions to the deep inelastic scattering
$\gamma^* p \to X p^\prime$ cross sections, which leave the target
intact, contribute at leading twist to deep inelastic scattering.
Diffractive events resolve the quark-gluon structure of the
virtual photon, not the quark-gluon structure of the target, and
thus they give contributions to structure functions which are not
target parton probabilities. Our analysis of deep inelastic
scattering $\gamma^*(q) p \to X$, when interpreted in frames with
$q^+ > 0,$ also supports the color dipole description of deep
inelastic lepton scattering at small $x_{bj}$.  For example, in
the case of the aligned-jet configurations, one can understand
$\sigma_T(\gamma^* p)$ at high energies as due to the coherent
color gauge interactions of the incoming quark-pair state of the
photon interacting, first coherently and finally incoherently, in
the target.

The distinction between structure functions
and target parton probabilities is also implied by the Glauber-Gribov
picture of nuclear
shadowing~\cite{Gribov:1969jf,Brodsky:1969iz,Brodsky:1990qz,Piller:2000wx}.
In this framework, shadowing arises from interference between
complex rescattering amplitudes involving on-shell intermediate
states.  In contrast, the wave function
of a stable target is strictly real since it does not have on
energy-shell configurations.  Thus nuclear shadowing is not a
property of the light-front wavefunctions of a nuclear target;
rather, it involves the total dynamics of the $\gamma^*$ - nucleus
collision.
A strictly probabilistic interpretation of the deep inelastic cross
section cross section is thus precluded.

\section{Acknowledgment} I thank Professor Harald Fritzsch and the Von
Humboldt Foundation for inviting me to the Heisenberg Symposium.  Much of
the work reported here was done in collaboration with others, especially,
Markus Diehl, John Hiller, Paul Hoyer, Dae
Sung Hwang, Nils Marchal, Bo-Qiang Ma, Gary McCartor,
Hans Christian Pauli, Stephane Peigne, Francesco Sannino, Ivan Schmidt,
and Prem Srivastava.  This work was supported by the Department of Energy
under contract number DE-AC03-76SF00515.


\begin{thebibliography}{00}


\bibitem{Dirac:cp}
P.~A.~Dirac,
Rev.\ Mod.\ Phys.\  {\bf 21}, 392 (1949).

\bibitem{Brodsky:1997de}
For a review and references, see S.~J.~Brodsky, H.~C.~Pauli and
S.~S.~Pinsky,
Phys.\ Rept.\  {\bf 301}, 299 (1998)
[arXiv:hep-ph/9705477].

\bibitem{Srivastava:2000cf}
P.~P.~Srivastava and S.~J.~Brodsky,
Phys.\ Rev.\ D {\bf 64}, 045006 (2001)
[arXiv:hep-ph/0011372].




\bibitem{Brodsky:2001ii}
S.~J.~Brodsky, D.~S.~Hwang, B.~Ma and I.~Schmidt,
Nucl.\ Phys.\ B {\bf 593}, 311 (2001) [hep-th/0003082].



\bibitem{Lepage:1980fj}
G.~P.~Lepage and S.~J.~Brodsky,
Phys.\ Rev.\ D {\bf 22}, 2157 (1980).

\bibitem{Maskawa:1975ky}
T.~Maskawa and K.~Yamawaki,
Method Of Quantization,''
Prog.\ Theor.\ Phys.\  {\bf 56}, 270 (1976).


\bibitem{Pauli:1985pv}
H.~C.~Pauli and S.~J.~Brodsky,
Phys.\ Rev.\ D {\bf 32}, 1993 (1985).

\bibitem{Abada:2001if}
A.~Abada, P.~Boucaud, G.~Herdoiza, J.~P.~Leroy, J.~Micheli,
O.~Pene and J.~Rodriguez-Quintero,
Phys.\ Rev.\ D {\bf 64}, 074511 (2001)
[arXiv:hep-ph/0105221].


\bibitem{Matsumura:1995kw}
Y.~Matsumura, N.~Sakai and T.~Sakai,
Phys.\ Rev.\ D {\bf 52}, 2446 (1995) [arXiv:hep-th/9504150].

\bibitem{Hiller:2001mh}
J.~R.~Hiller, S.~Pinsky and U.~Trittmann,
supersymmetric Yang-Mills theory,''
Phys.\ Rev.\ D {\bf 64}, 105027 (2001)
[arXiv:hep-th/0106193].

\bibitem{Brodsky:2001ja}
S.~J.~Brodsky, J.~R.~Hiller and G.~McCartor,
hep-ph/0107038.


\bibitem{Bardeen:1979xx}
W.~A.~Bardeen, R.~B.~Pearson and E.~Rabinovici,
Phys.\ Rev.\ D {\bf 21}, 1037 (1980).

\bibitem{Dalley:2001gj}
S.~Dalley,
Phys.\ Rev.\ D {\bf 64}, 036006 (2001)
[arXiv:hep-ph/0101318].


\bibitem{Burkardt:2001dy}
M.~Burkardt and S.~K.~Seal,
arXiv:hep-ph/0105109.


\bibitem{Pauli:2001vi}
H.~C.~Pauli,
arXiv:hep-ph/0111040.


\bibitem{McCartor:hj}
G.~McCartor,
in Proc. of  New Nonperturbative Methods and Quantization
of the Light Cone, Les Houches, France, 24 Feb - 7 Mar 1997.


\bibitem{Yamawaki:1998cy} K.~Yamawaki,
arXiv:hep-th/9802037.



\bibitem{BH2}
S.~J.~Brodsky and D.-S. Hwang, in preparation.


\bibitem{Brodsky:1973kb}
S.~J.~Brodsky, R.~Roskies and R.~Suaya,
Momentum Frame,''
Phys.\ Rev.\ D {\bf 8}, 4574 (1973).


\bibitem{Cornwall:1989gv}
J.~M.~Cornwall and J.~Papavassiliou,
Phys.\ Rev.\ D {\bf 40}, 3474 (1989).



\bibitem{Brodsky:1994eh}
S.~J.~Brodsky and H.~J.~Lu,
Phys.\ Rev.\ D {\bf 51}, 3652 (1995)
[arXiv:hep-ph/9405218].


\bibitem{Brodsky:2001dx}
S.~J.~Brodsky,
arXiv:hep-ph/0102051.

\bibitem{Brodsky:1980zm}
S.~J.~Brodsky and S.~D.~Drell,
Phys.\ Rev.\ D {\bf 22}, 2236 (1980).


\bibitem{Brodsky:2000xy}
S.~J.~Brodsky, M.~Diehl and D.~S.~Hwang,
Nucl.\ Phys.\ B {\bf 596}, 99 (2001)
[arXiv:hep-ph/0009254].


\bibitem{Diehl:2000xz}
M.~Diehl, T.~Feldmann, R.~Jakob and P.~Kroll,
Nucl.\ Phys.\ B {\bf 596}, 33 (2001)
[Erratum-ibid.\ B {\bf 605}, 647 (2001)]
[arXiv:hep-ph/0009255].



\bibitem{Brodsky:1999hn}
S.~J.~Brodsky and D.~S.~Hwang,
Nucl.\ Phys.\ B {\bf 543}, 239 (1999) [hep-ph/9806358].


\bibitem{Okun}
L.~Okun and  I.~Yu.~Kobsarev, ZhETF,
\textbf{ 43} (1962) 1904 (English
translation: JETP \textbf{ 16} (1963) 1343);
L.~Okun, in proceedings of the 4th
International Conference on Elementary Particles, Heidelberg,
Germany (1967), edited by H.~Filthuth, North-Holland (1968).

\bibitem{Kob70}
I.~Yu.~Kobsarev and V.~I.~Zakharov, Ann. Phys. {\bf 60}
(1970) 448.

\bibitem{Teryaev}
O. V. Teryaev, hep-ph/9904376 (1999).


\bibitem{Brodsky:2001ue}
S.~J.~Brodsky, P.~Hoyer, N.~Marchal, S.~Peigne and F.~Sannino,
hep-ph/0104291.

\bibitem{Leibbrandt:1987qv}
G.~Leibbrandt,
Rev.\ Mod.\ Phys.\  {\bf 59}, 1067 (1987).

\bibitem{Gribov:1969jf}
V.~N.~Gribov,
Sov.\ Phys.\ JETP {\bf 29}, 483 (1969) [Zh.\ Eksp.\ Teor.\ Fiz.\
{\bf 56}, 892 (1969)].


\bibitem{Brodsky:1969iz}
S.~J.~Brodsky and J.~Pumplin,
Phys.\ Rev.\  {\bf 182}, 1794 (1969).


\bibitem{Brodsky:1990qz}
S.~J.~Brodsky and H.~J.~Lu,
Phys.\ Rev.\ Lett.\  {\bf 64}, 1342 (1990).




\bibitem{Piller:2000wx}
G.~Piller and W.~Weise,
Phys.\ Rept.\  {\bf 330}, 1 (2000) [hep-ph/9908230].







\end{thebibliography}
\end{document}